\documentclass[%
 reprint,%
 amssymb, amsmath,%
 aip,apl,
]{revtex4-1}

\usepackage{docs}%
\usepackage{bm}%

\usepackage{graphicx}

\begin{document}

\title{Determination of optimal reversed field with maximal electrocaloric cooling by a direct entropy analysis}%

\affiliation{Institute of Materials Science, Technische Universit{\"a}t Darmstadt, 64287 Darmstadt, Germany}
\author{Yang-Bin Ma}

\email{y.ma@mfm.tu-darmstadt.de}

\author{Nikola Novak}

\author{Karsten Albe}

\author{Bai-Xiang Xu}
\email{xu@mfm.tu-darmstadt.de}
\date{\today}%

\begin{abstract}
Application of a negative field on a positively poled ferroelectric sample can enhance the electrocaloric cooling and appears as a promising method to optimize the electrocaloric cycle. Experimental measurements show that the maximal cooling does not appear at the zero-polarization point, but around the shoulder of the P-E loop. This phenomenon cannot be explained by the theory based on the constant total entropy assumption under adiabatic condition. In fact, adiabatic condition does not imply constant total entropy when irreversibility is involved. A direct entropy analysis approach based on work loss is proposed in this work, which takes the entropy contribution of the irreversible process into account. The optimal reversed field determined by this approach agrees with the experimental observations. This study signifies the importance of considering the irreversible process in the electrocaloric cycles.
\end{abstract}

\maketitle

\begin{figure} [htp]
 \centering 
 \centerline{\includegraphics[width=8.5cm]{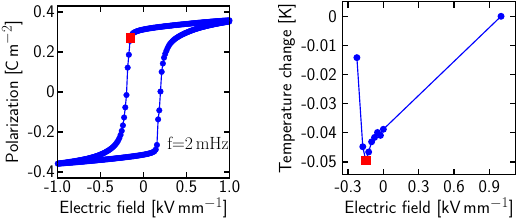}}
 \caption{Direct EC measurement on single crystals of Pb(Mg$_{1/3}$/Nb$_{2/3}$)$_{0.71}$Ti$_{0.29}$O$_3$ (PMN-29PT) at 303.0\,K.}
 \label{fig:figure1}
\end{figure}

The electrocaloric (EC) effect shows great application potential in the technology of solid state refrigeration.~\cite{2009_Kar-Narayan,2012_Jia,2014_Gu}
Even though much effort has been made to explore material candidates with large EC effect and device concepts, there are few work concerning the optimization of electrocaloric cycle. 
In a conventional EC cycle of a solid refrigerant, the cooling effect is obtained simply by removing the previously applied electric field. 
Using direct heat flux calorimetry on poly(vinylidene fuoridetride-trifuoroethylene) films, Basso {\em et al.}~\cite{2014_Basso} demonstrated that the electrocaloric cooling can be doubled if a negative electric field to a positively poled sample is applied. The EC hysteresis of ferroelectric ceramics measured by Thacher {\em et al.}~\cite{1968_Thacher} indicated also that a reversed electric field can increase the cooling effect of ferroelectric ceramics. In the authors' previous work,~\cite{2016_Ma} experimental and numerical studies were carried in PMN-29PT and BaTiO$_3$, which demonstrated that there exists an optimal reversed electric field, corresponding to a position around the shoulder of the dielectric hysteresis. At this point, the EC cooling effect reaches its maximum (also see Fig.~\ref{fig:figure1}). This phenomenon was also observed in Ref.~\onlinecite{2014_Basso}.
It is of scientific and engineering importance to determine and understand this optimal reversed electric field. In the literature, as reviewed in Ref.~\onlinecite{2015_Kutnjak}, the EC cycle is considered to be reversible, with constant total entropy under the adiabatic condition. This assumption leads to the conclusion that the maximal cooling takes place at the zero polarization point, since at this point the dipolar entropy takes maximum (see Model I). This conclusion deviates obviously from the experimental observations. In fact, the total entropy $S_\mathrm{total}$ should satisfy:
\begin{equation} \label{eq:dip-par}
{\Delta S_\mathrm{total} = \Delta S_\mathrm{dip} + \Delta S_\mathrm{vib}} 
 \begin{cases}
  = 0 
  & \text{in reversible process},
 \\
  \neq 0 
  & \text{in irreversible process}, \nonumber
 \end{cases}
\end{equation}
where $S_\mathrm{dip}$ and $S_\mathrm{vib}$ are the dipolar and the vibrational entropy, respectively. In the case of applying a reversed electric field, the irreversible contribution becomes considerable. In other words, to correctly determine the optimal reversed electric field, the change of the total entropy induced by the work loss $W_\mathrm{loss}$ due to the irreversible process should be evaluated.

The Maxwell relation is valid under the assumption that the process is thermodynamically reversible. Hence, it is inappropriate to utilize the indirect approach~\cite{2012_Zhang,2008_Akcay,2016_Marathe} based on the Maxwell relation for the issue of interest.
The adiabatic condition can be strictly fulfilled in Monte-Carlo and Molecular Dynamics simulations by utilizing the direct method.~\cite{2012_Ponomareva,2012_Rose,2015_Ma,2016_Ma,2016_Marathe,2016_Gruenebohm} In these methods, the information of the entropy change can be hardly extracted, which is, however, very important for the understanding of the maximal EC cooling point. In this study, we propose a direct method to analyze the entropy changes in an irreversible process, and apply it to reveal the optimal reversed electric field which maximizes the EC cooling.  

According to Ref.~\onlinecite{2015_Kutnjak}, the dipolar mean field free energy $F_\mathrm{dip}$ can be expressed as
\begin{eqnarray} \label{eq:mean}
F_\mathrm{dip} = F_0 + ( \frac{1}{2} a P^2 + \frac{1}{4} b P^4 + \frac{1}{6} c P^6) - E P,
\end{eqnarray}
where $F_0$ is the field-independent part, the phenomenological coefficients $a$, $b$, $c$ should be temperature-dependent, and $P$ is the total macroscopic polarization. For simplicity, the sixth order-term and the temperature-dependence of $b$ are abandoned while $ a = a_0 (T_\mathrm{A} - T_\mathrm{C}) $ with $T_\mathrm{A}$ and $T_\mathrm{C}$ as the initial temperature and the Curie temperature, respectively. If not specified, the initial temperature $T_\mathrm{A}$ is set to be 0.7. 
According to Pirc and Blinc~\cite{1999_Pirc}, the following normalized parameters are chosen: $a_0=1$, $T_\mathrm{C}= 1$, $ b = 1/3 $, and $ a_0 = \partial a / \partial T = 1$.
The dipolar entropy $S_\mathrm{dip}$ can be derived from as 
$
{ S_\mathrm{dip} } = -\partial F_\mathrm{dip} /\partial T_\mathrm{A} = - \frac{1}{2} a_0 P^2 
$. Therefore the change of $S_\mathrm{dip}$ from state $\mathrm{A}$ with polarization $P_\mathrm{A}$ to another state with polarization $P$ can be simply expressed as 
\begin{equation}
\label{eq:deltasconf}
 { \Delta S_\mathrm{dip} } = - \frac{1}{2} a_0 {P}^2 + \frac{1}{2} a_0 {P_\mathrm{A}}^2.
\end{equation} 

The change of $S_\mathrm{vib}$, can be approximated as~\cite{2015_Kutnjak}
\begin{eqnarray} \label{eq:deltasvib}
 { \Delta S_\mathrm{vib} }
 &=& \int_{T_\mathrm{A}}^{T} \frac{C_\mathrm{ph}}{T} dT 
 \cong C_\mathrm{ph} ln(T / T_\mathrm{A}),
\end{eqnarray}
where $T$ is the current temperature, and $C_\mathrm{ph}$ is the specific heat capacity of the non-polar degrees of freedom, which can be assumed to be temperature-independent~\cite{2015_Kutnjak}. Based on the work~\cite{1999_Pirc}, the normalized value $C_\mathrm{ph} = 15$ is applied for the study in this work.

In the following three different models are applied to determine the optimal magnitude of the reversed field. 
\begin{figure} [htp]
 \centering 
 \centerline{\includegraphics[width=5.5cm]{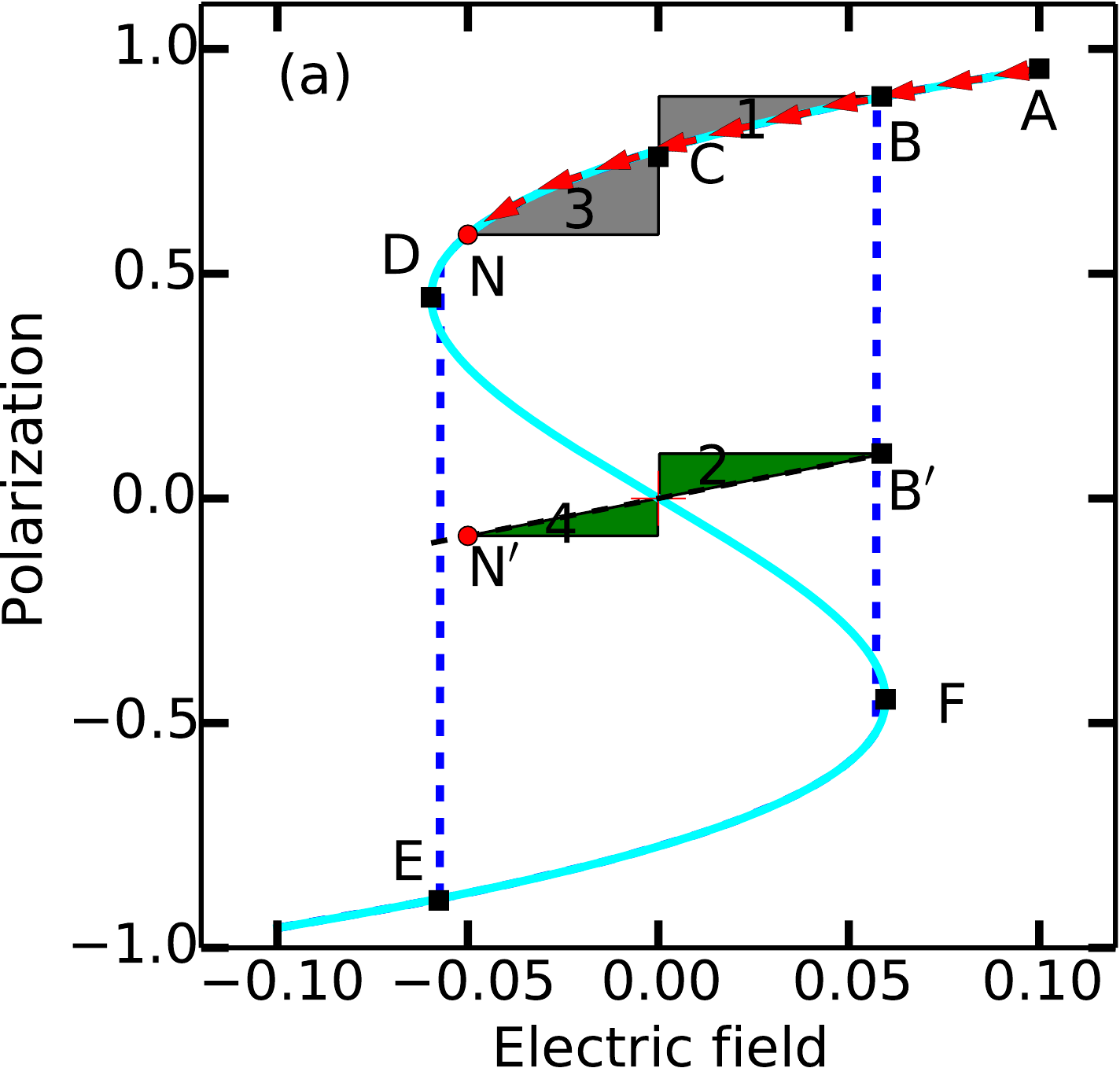}}
 \caption{Decomposition of the polarization into reversible and irreversible contributions.
   }
  \label{fig:figure2}
\end{figure}
\begin{figure*} [htp]
 \centering 
 \centerline{\includegraphics[width=15cm]{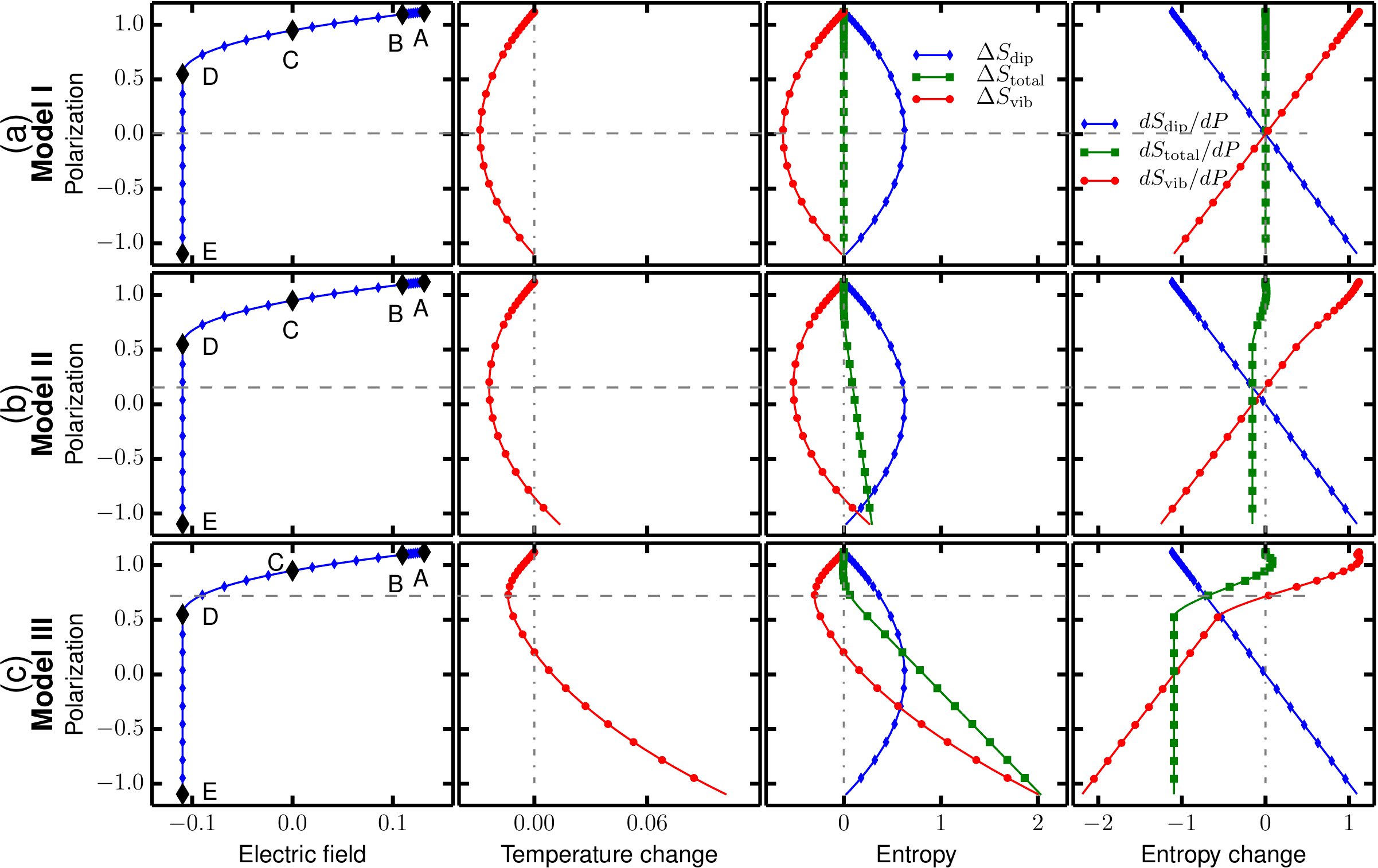}}
 \caption{Determination of the optimal reversed field with maximal cooling. (a), (b) and (c) are based on Model I, II, and III, respectively. Gray horizontal lines indicate the positions with maximal cooling, which are always the points where $ d S_\mathrm{total} / d P = d S_\mathrm{dip} / d P $ for all models.  }
\label{fig:figure3}
\end{figure*}
\\[1ex]
\noindent\textbf{Model I.} Assuming a reversible process, adiabatic condition implies $\Delta S_\mathrm{total}=0$, and thus $\Delta S_\mathrm{vib}=-\Delta S_\mathrm{dip}$. Application of Eq.~(\ref{eq:deltasconf}) and ~(\ref{eq:deltasvib}) leads to
\begin{eqnarray} \label{eq:sol1}
T &=& T_\mathrm{A} \exp \big[ ( \frac{1}{2} a_0 {P}^2 - \frac{1}{2} a_0 {P_\mathrm{A}}^2 ) / C_\mathrm{ph} \big].
\end{eqnarray}
This result has been obtained in e.g. Ref.~\onlinecite{2015_Kutnjak}. For the state between A and E, one can determine the corresponding temperature change by Eq.~(\ref{eq:sol1}), the entropy changes, and the derivatives of these changes with respect to the polarization. Results are shown in Fig.~\ref{fig:figure3}(a). It is seen that the maximal temperature drop happens at the point $P=0$. This is due to the fact that Eq.~(\ref{eq:sol1}) depends on $P^2$. When the polarization equals to 0, there appears the highest disorder of polarization, i.e., a maximum of $S_\mathrm{dip}$ and a corresponding minimum of $S_\mathrm{vib}$ since the total entropy is constant. In other words, ignoring the irreversible contribution leads to the conclusion that the maximum EC cooling appears at $P=0$. This deviates from the previous experimental~\cite{2014_Basso,2016_Ma} and numerical observation~\cite{2016_Ma}. 
\\[1ex]
\noindent\textbf{Model II.} Considering the irreversible contribution, the total entropy is related to the change of the work loss as
\begin{equation}
\label{eq:St}
{ d S _ \mathrm{total} } = \frac{d W_\mathrm{loss}}{ T } \approx \frac{d W_\mathrm{loss}}{ T_\mathrm{A} } .
\end{equation} 
Note that the approximation in the last equation results from the fact that the temperature change is rather small compared with the initial temperature $T_\mathrm{A}$.
Then through integration Eq.~(\ref{eq:St}) can be rewritten as 
\begin{eqnarray} \label{eq:eq6}
{ \Delta S _ \mathrm{total} } &=& \frac{ W_\mathrm{loss}}{ T_\mathrm{A} } = \Delta S_\mathrm{dip} + \Delta S_\mathrm{vib}.
\end{eqnarray}
Insertion of Eq.~(\ref{eq:deltasconf}) and ~(\ref{eq:deltasvib}) into Eq.~(\ref{eq:eq6}), one has
\begin{eqnarray} \label{eq:sol2}
T &=& T_\mathrm{A} \exp \big[ ( W_\mathrm{loss} / T_\mathrm{A} - \Delta S_\mathrm{dip} ) / C_\mathrm{ph} \big].
\end{eqnarray}
Eq.~(\ref{eq:sol2}) is reduced to Eq.~(\ref{eq:sol1}) if the process is reversible with $W_\mathrm{loss} = 0$.

Next we show how $W_\mathrm{loss}$ can be evaluated based on the decomposition of the polarization and the Landau theory. Setting the first derivative of the Landau free energy $\partial F_\mathrm{dip} / \partial P $ to be zero, one has the corresponding polarization for the actual electric field $E$ from
\begin{eqnarray} 
 { E } &=& a P + b { P }^3.
 \label{eq:act}
\end{eqnarray}
This is demonstrated in the S-shaped curve in the P-E plane below the Curie temperature, as it is shown in Fig.~\ref{fig:figure2}. Through the snap-through construction D-E and F-B, one obtains the ideal P-E loop. Take the saturated point A as the initial state with polarization $P_\mathrm{A}$, initial temperature $T_\mathrm{A}$ and the electric field $E_\mathrm{A}$. Denote by C the point with zero electric field, and by D the inflection point of the S-shaped curve. Between A and B the process is fully reversible, while between B and E irreversibility is involved. 
Bolten {\em et al.}~\cite{2000_Bolten} pointed out that in the ferroelectric P-E loops, the reversible part can be described by a straight line without any hysteretic heat loss since the contribution mainly arises from the ionic and electronic polarization. This straight line should cross through the center of the hysteresis.
In this work, we use the tangent line at the point B as the slope of the reversible part, since at point B irreversible part starts to be involved.
The slope is $ \frac{\partial E}{\partial P} \mid _ { P=P_\mathrm{B} } = a + 3 b {P_\mathrm{B}} ^2 $, where $ E_\mathrm{B} = a P_\mathrm{B} + b P_\mathrm{B}^3$.
Then the reversible polarization $P_{\mathrm{r}}$ is given in
\begin{eqnarray} 
E = (a + 3 b {P_\mathrm{B}} ^2 ) P_{\mathrm{r}} .
\label{eq:rev}
\end{eqnarray}
From Eq.~(\ref{eq:act}) and (\ref{eq:rev}), it follows
\begin{eqnarray} \label{eq:work}
 { a P + b P ^3 } &=& (a + 3 b {P_\mathrm{B}} ^2 ) P_{\mathrm{r}}.
\end{eqnarray}
In this way the reversible part of the polarization $P$ can be expressed as
\begin{eqnarray}
{ P_{\mathrm{r}} (P) } &=& 
\frac{ a P + b P^3 }{ a + 3 b { P_{ \mathrm{B} } } ^2 }.
\end{eqnarray}
The work loss $W_\mathrm{loss}$ is termed as the difference between the work done in the actual process and that in the reversible process under the same electric field $ E$. For any point located between B and D, the work loss is given through the difference of two integrals:
\begin{eqnarray} 
& & { W_\mathrm{loss1} (P) }  = W_\mathrm{actual} - W_\mathrm{r} \nonumber \\
& & =\int_{P_\mathrm{B}}^{P} E dP  - \int_{P_\mathrm{B^\prime}}^{P_\mathrm{r}} E_\mathrm{r} dP_{\mathrm{r}}  \nonumber \\
& & = \frac{1}{2} a { P }^2 + \frac{1}{4} b { P }^4 
  - ( \frac{1}{2} a { P_{\mathrm{B}} }^2 + \frac{1}{4} b { P_{\mathrm{B}} }^4 )  \nonumber \\ 
  & & - \frac{1}{2} a { P_\mathrm{r} }^2 - \frac{3}{2} b { P_{\mathrm{B}} }^2 { P_\mathrm{r} }^2 + \frac{1}{2} a { P_\mathrm{B^\prime}}^2 + \frac{3}{2}b { P_{\mathrm{B}} }^2 { P_{\mathrm{B^\prime}} }^2  
 \label{eq:lost}
 \end{eqnarray}
where $P_\mathrm{B^\prime}=P_\mathrm{r}|_{P=P_{\mathrm{B}}} = \frac{ a P_\mathrm{B} + b P_\mathrm{B}^3 }{ a + 3 b { P_{ \mathrm{B} } } ^2 } $. For the point N illustrated in Fig.~\ref{fig:figure2}, the work loss Eq.~(\ref{eq:lost}) can be equivalently evaluated from the area difference, more exactly  $( V_2-V_1 )+ (V_3 - V_4 )$ with $V_i$ being the area of the region $i$ shown in Fig.~\ref{fig:figure2}. 
For any point between D and E, one has to go through the vertical snap-through line, which represents the hysteretic path. It is not difficult to obtain the work loss in the following form
\begin{eqnarray} 
{ W_\mathrm{loss2} (P) } &=& W_\mathrm{loss1}|_{P=P_\mathrm{D}} + (P - P_{\mathrm{D}}) E_{\mathrm{D}}.
 \label{eq:lost2}
\end{eqnarray}
In summary, 
\begin{equation} 
{W_\mathrm{loss}} = 
 \begin{cases}
  W_\mathrm{loss1} 
  & \text{for points between B and D },
 \\
  W_\mathrm{loss2} 
  & \text{for points between D and E }. \nonumber
 \end{cases}
\end{equation}
After $W_\mathrm{loss}$ is determined through Eq.~(\ref{eq:lost}) or (\ref{eq:lost2}), the temperature can be calculated by Eq.~(\ref{eq:sol2}) for the irreversible process. 

The results based on this model are presented in Fig.~\ref{fig:figure3}(b). Different from model I, the maximum EC cooling takes place at a position before the polarization vanishes (see the gray dashed line). The underlying reason is due to the entropy contribution of the irreversible process.
From B to C, it is an almost reversible process, and this brings about a steady decrease of the temperature. 
Within the region C-D $W_\mathrm{loss}$ and $\Delta S_\mathrm{total}$ increase steadily with the decrease of the polarization, which leads to a weaker decrease of the temperature compared to the strong decrease within A-C.
In the region D-E the irreversible process dominates, and thus $W_\mathrm{loss}$ and $\Delta S_\mathrm{total}$ increase significantly.
During polarization decreases in this region, the increase of $ S_\mathrm{total}$ gradually overtakes that of $S_\mathrm{dip} $, which makes the tendency of temperature decrease weaker.
Finally, at the point where $d S_\mathrm{total} / d P$ is equal to $d S_\mathrm{dip} / d P$, a maximum EC cooling effect is reached (see the gray dashed line).
Afterwards, $d S_\mathrm{total}/ d P$ becomes larger than $d S_\mathrm{dip}/ d P$, and the cooling effect becomes weaker and can even be switched to heating.

It demonstrates that the entropy contribution of the irreversible process shifts the position with maximal EC effect to a state with positive polarization. However, the shift is still far away from point D, determined from the experimental results.  
\\[1ex]
\noindent\textbf{Model III.}
As temperature increases and approaches $T_\mathrm{C}$, the hysteresis becomes slimmer, and the polarization switching is more moderate around the coercive field. The ideal scheme of the P-E loop considered in the model II, where at the coercive field the polarization is assumed to switch immediately from one direction to another direction, is inappropriate. In order to consider the P-E loop in a more realistic scheme, a temperature-dependent factor $\alpha$ is introduced to correct $ W_\mathrm{loss}$ in Eq.~(\ref{eq:sol2}),
\begin{eqnarray} \label{eq:sol3}
T &=& T_\mathrm{A} \exp \big[ ( \alpha W_\mathrm{loss} / T_\mathrm{A} - \Delta S_\mathrm{dip} ) / C_\mathrm{ph} \big],
\end{eqnarray}
where $\alpha$ is the factor between the work loss obtained considering a non-ideal P-E loop and the one using the model II. Since in the realistic case, the work loss is larger than that evaluated in the model II, the factor $\alpha > 1$. Meanwhile, $\alpha$ increases with temperature.
However, the exact relation between $\alpha$ and the temperature is not available in the literature. 
Nonetheless, it is known that at higher temperature the polarization switching within the metastable state by the Landau approach deviates further from that in realistic materials.
In other words, the work loss is underestimated more within the metastable state at higher temperatures, if Eq.~(\ref{eq:sol2}) is utilized.
The extent of the underestimation is assumed to be related to the radius of the curvature of S-shaped curve at point D ${R_\mathrm{D}}$. 
From the Landau free energy Eq.~(\ref{eq:mean}), one has $R_\mathrm{D} = 1 / \sqrt{- 12 a b}$. At 0 K, the radius of the curvature at point D is then $R_0 = 1 / \sqrt{12 a_0 T_\mathrm{C} b}$. Since the work loss is related to the area circumvented in the hysteresis, it is assumed that
$\alpha$ is proportional to normalized ${R_\mathrm{D}}^2$ in the following fashion
\begin{eqnarray} \label{eq:alpha}
\alpha = \lambda {R_\mathrm{D}}^2 / {R_0}^2 = \lambda T_\mathrm{C} / ( T_\mathrm{C} - T_\mathrm{A} ).
\end{eqnarray}
The factor $\lambda$ should be independent of temperature and reflects the ratio of the work loss due to other materials-complexity, e.g. the switchable defects and the complex domain wall movements. Hereby $\lambda$ is set to be 2.0.
 
The corresponding results on temperature change and entropy changes are depicted in Fig.~\ref{fig:figure3}(c). In comparison with the model II, the work loss in the model III is corrected through consideration of the extra contribution. Hence, the maximum EC cooling point is further shifted away from the zero-polarization point, and reaches around the point D, showing good agreement with the experimental results~\cite{2014_Basso,2016_Ma}. The exact position is also defined as the point where $d S_\mathrm{total} / d P$ catches up with $d S_\mathrm{dip} / d P$ (see the gray dashed line). The explanation provided in the model II remains valid here. \\[2ex]

In conclusion, the entropy contribution of irreversible process plays an important role in the study of EC effect. Adiabatic condition does not necessarily imply constant total entropy. In the particular case of negative EC effect induced by reversed electric field, we show that only when the work loss due to the irreversible process is appropriately considered, the optimal reversed field with maximal cooling effect can be accurately determined.

Results based on the model III (not shown) also demonstrate that with increasing initial temperature, the optimal electric field with the maximum EC cooling decreases, and closer to the coercive field. The maximal EC cooling also decreases. 

The funding of Deutsche Forschungsgemeinschaft (DFG) SPP1599 (XU 121/1-2, AL 578/16-2, NO 1221/2-1) is gratefully acknowledged.

\end{document}